\documentclass[pra,twocolumn,showpacs,preprintnumbers,amsmath,amssymb,superscriptaddress]{revtex4}

\usepackage[pdftex]{graphicx}
\usepackage{dcolumn}
\usepackage{bm}

\usepackage[T1]{fontenc}
\usepackage[latin1]{inputenc}
\usepackage[english]{babel}
\usepackage{amsmath}
\usepackage{amssymb}
\usepackage{amstext}
\usepackage{braket}
\usepackage{relsize}
\usepackage{subfigure}

\newcommand{\ud}{\mathrm{d}}
\newcommand{\ue}{\mathrm{e}}

\newcommand{\comment}[1]{} 

\begin{document}

\preprint{}

\title{Complete devil's staircase and crystal--superfluid transitions in a dipolar XXZ spin chain: A trapped ion quantum simulation}

\author{Philipp Hauke}
\email{Philipp.Hauke@icfo.es}
\author{Fernando M. Cucchietti}
\affiliation{
ICFO -- Institut de Ci\`encies Fot\`oniques, Mediterranean Technology Park, E-08860 Castelldefels (Barcelona), Spain
}
\author{Alexander M\"{u}ller-Hermes}
\author{Mari-Carmen Ba\~{n}uls}
\author{J. Ignacio Cirac}
\affiliation{Max-Planck-Institut f\"ur Quantenoptik, Hans-Kopfermann-Str.\ 1, D-85748 Garching, Germany}

\author{Maciej Lewenstein}
\affiliation{
ICFO -- Institut de Ci\`encies Fot\`oniques, Mediterranean Technology Park, E-08860 Castelldefels (Barcelona), Spain
}
\affiliation{
ICREA -- Instituci\`{o} Catalana de Ricerca i Estudis Avan\c{c}ats, E-08010 Barcelona, Spain
}

\date{\today}

\begin{abstract}

Systems with long-range interactions show a variety of intriguing properties: 
they typically accommodate many meta-stable states, 
they can give rise to spontaneous formation of supersolids, and they can lead to 
counterintuitive thermodynamic behavior. 
However, the increased complexity that comes with long-range interactions strongly hinders theoretical studies.
This makes a quantum simulator for long-range models highly desirable.
Here, we show that a chain of trapped ions can be used to quantum simulate a one-dimensional model of hard-core bosons 
with dipolar off-site interaction and tunneling, equivalent to a dipolar XXZ spin-1/2 chain. 
We explore the rich phase diagram of this model in detail, employing perturbative mean-field 
theory, exact diagonalization, and quasi-exact numerical techniques 
(density-matrix renormalization group and infinite time evolving block decimation). 
We find that the complete devil's staircase --- an infinite sequence of crystal states existing at vanishing tunneling --- 
spreads to a succession of lobes similar to the Mott-lobes found in Bose--Hubbard models. 
Investigating the melting of these crystal states at increased tunneling, we do not find (contrary to similar two-dimensional models)
clear indications of supersolid behavior in the region around the melting transition.
However, we find that inside the insulating lobes there are quasi-long range (algebraic) correlations, opposed to 
models with nearest-neighbor tunneling which show exponential decay of correlations.

\end{abstract}

                          
\maketitle


\begin{section}{Introduction}

Dipolar interactions in a lattice system can lead to exciting new physics not present in systems 
with short-range interactions \cite{Lahaye2009}.
For example, they can give rise to new ground states, like crystal states with fractional filling factors or supersolids.
For instance, on the square lattice under hole-doping from half-filling, dipolar interactions can stabilize a supersolid phase \cite{Capogrosso2010}, in contrast to interactions of limited range \cite{Batrouni1995,Sengupta2005,Dang2008}.
Further, dipolar interactions can lead to the appearance of a large number of metastable states \cite{Menotti2007b}.
On the one hand, these can complicate finding the ground state in numerical studies, while on the other hand they might be useful for applications in quantum information and as quantum memories \cite{Trefzger2008,Braungardt2007,Pons2007}.
Systems with strong long-range interactions (where the integral over the interactions does not converge) can even exhibit counterintuitive thermodynamic effects like super-extensive behavior, breaking of ergodicity, or a non-concave entropy \cite{Mukamel2009}.
These effects, however, are not the subject of the present work, since we consider weak long-range interactions.

Since long-range interactions can make theoretical calculations more difficult, it would be highly desirable to study all this extremely rich physics with an analog quantum simulator
-- specially for dynamics or two or more dimensions. In one dimension (which is the subject of this work), standard analytical and numerical techniques typically suffice to give a good picture of the phase diagram, but this is an important limit to gain intuition. Its experimental implementation allows to judge the reliability and performance of analogue quantum simulators.
Natural candidate systems for this type of quantum simulation are trapped ions, 
where the high degree of control over state preparation, evolution, and readout allows access to a wide range of models.
In 2004, trapped ions were first proposed for quantum simulations of certain classes of spin models by Porras and Cirac \cite{Porras2004a}.
In a different context, similar spin models were derived earlier by Mintert and Wunderlich \cite{Mintert2001}, 
who studied the possibility of individual ion addressing using inhomogeneous magnetic fields. 
Recent proof-of-principle experiments showed that, indeed, quantum spin systems can be 
simulated faithfully by systems of cold trapped ions \cite{Friedenauer2008,Kim2010}.
For a review of trapped ion quantum simulations see \cite{Johanning2009}.

An advantage of trapped ions is that dipolar interactions (which are the most common 
type of long-range interactions) are naturally achieved without additional experimental effort 
\cite{Porras2004a}, since the interactions are mediated by phonons. 
This leads also to the peculiarity that, contrary to other possible setups for the quantum simulation of spin models, 
not only two-body (density--density) interactions, but also tunneling terms can be made long ranged.

In this work, we consider an experimentally feasible setup which allows to quantum 
simulate a chain of XXZ spins with dipolar interactions in a magnetic field. 
The system, which, as we show, supports a rich ground-state phase diagram, is described by the Hamiltonian
\begin{eqnarray}
\label{H}
    H &=& J \sum_{i<j} \frac{1}{\left|i-j\right|^3} 
    								\left[ \cos\theta S_i^{\,z} S_j^{\,z} + 
    											 \sin\theta \left(  S_i^{\,x} S_j^{\,x} + S_i^{\,y} S_j^{\,y} \right) \right] \nonumber \\
    	 & &	-\mu \sum_i S_i^{\,z}\,,
\end{eqnarray}       	 
where $\mu$ is the chemical potential (equivalent to an external magnetic field), and the $S_i^{\,\alpha}$ are spin-operators at site $i$.

The problem can be mapped to a system of hard-core bosons using the standard Holstein--Primakoff transformation, $S_i^z\to n_i-1/2$, $S_i^+\to a_i^\dagger$, and $S_i^-\to a_i$, where $a_i^\dagger$ ($a_i$) creates (destroys) a boson at site $i$ and where $n_i$ denotes the corresponding number operator. 
In this picture, an up-spin corresponds to an occupied site and a down-spin to an empty one, the ZZ terms translate to dipolar off-site density-density interactions, and the XX and YY interactions become long-range tunneling terms. Since typically tunneling terms are short-ranged, the latter has to be seen as an important peculiarity of our model.
In the following, we will use both the spin and boson pictures interchangeably, 
because some aspects are better described in terms of spins, while others are more familiar in the language of hard-core bosons. 

By varying the angle $\theta$, we can explore all ranges of relative strength of the interactions, 
\emph{e.g.}\ the XX chain in a transverse field ($\theta=\pi/2$), or the Ising model ($\theta=0$) 
--- but both with long-range interaction.
Moreover, the model can be tuned from negative to positive XY interaction,
with the latter leading to a deformation of the phase diagram due to frustration effects.

At $\theta=0$, where Hamiltonian~\eqref{H} behaves essentially classical, this model is known to show a succession of insulating states with different rational filling factors \cite{Bak1982}.
This succession, called `devil's staircase', covers the entire range of the chemical potential, 
with (for an infinite system) each rational filling factor occuring on a finite range of chemical potential. 
Since the Hamiltonian \eqref{H} shows a symmetry of up-down spins (or in hard-core boson terminology, a particle-hole symmetry), 
we consider in this work only negative magnetization  (filling factors lower than $1/2$).
When $\left|\theta\right|$ grows, the insulating states become destabilized and finally melt to a superfluid phase.
We concentrate on characterizing the complete phase diagram of the model, and the nature of the
correlations in each phase. We find that the long-range nature of the interactions changes qualitatively
one important aspect of the physics of the problem: inside the insulating phases, the off-diagonal
correlations decay algebraically instead of exponentially (as is typical in models with short-range tunneling) \cite{Deng2005}.
We call this phase a quasi-supersolid, since this is the closest analogue to a supersolid that can exist in one dimension: 
A supersolid is characterized by the coexistence of diagonal and off-diagonal long-range order (LRO). In 1D, however, off-diagonal LRO cannot occur \cite{Mermin1966}, and the `best' that can exist is the coexistence of diagonal LRO and off-diagonal quasi-LRO with algebraically decaying correlations.
This quasi-supersolid is exceptional, since normally systems with dipolar interactions in 1D can be described by Luttinger theory \cite{Citro2007,Citro2008}. In that case, diagonal correlations decay with the Luttinger parameter $K$, and off-diagonal correlations with $1/K$. This means that if one of them decays slowly, the other one decays very fast. In contrast, in our model with long-range tunneling both diagonal and off-diagonal correlations decay slowly, and Luttinger theory is not a valid description. 

Recently, a similar model with short-range tunneling has been investigated by a strong-coupling expansion in the context of ultracold 
dipolar atomic gases \cite{Burnell2009b}.
Placed in a one-dimensional optical lattice, in the limit of strong on-site interaction, such a system becomes similar to the one described by 
Hamiltonian~\eqref{H}, but with nearest-neighbor (NN) instead of long-ranged tunneling. Hence, the quasi-supersolid phase cannot be observed in this system.
Throughout the paper, we draw comparisons between our model and the model with NN tunneling and dipolar interaction, as well as with a model with both NN tunneling and interaction (the NN-XXZ model).

We organize this paper as follows:
First, we briefly explain in Sec.~\ref{sec:experimentalimplementation} how the desired model Hamiltonian $H$ 
can be implemented in a system of trapped ions.
Then, in Sec.~\ref{sec:knownfacts}, we outline our expectations by discussing previous related results.
The following sections are dedicated to a thorough investigation of the ground state phase diagram of the model.
Our most accurate analysis of the problem comes from numerical density matrix renormalization group (DMRG), Sec.~\ref{sec:DMRG}. However, it is good to first form intuition via analytical approaches, even if approximate. We will present mean field and perturbative results in Sec.~\ref{sec:MF} to find the upper borders of the crystal lobes. Some limitations of DMRG can be overcome with other numerical techniques. For instance, we can study infinite systems (as opposed to finite sized) with the infinite time evolving block decimation (iTEBD) algorithm (section Sec.~\ref{sec:iTEBD}). Finally, experimentally relevant small systems can be thoroughly studied with exact diagonalization (Sec.~\ref{sec:ED}).
In Sec.~\ref{sec:conclusion}, we offer some final remarks.

\end{section}


\begin{section}{Experimental implementation \label{sec:experimentalimplementation}}

In this section, we shortly explain how the desired Hamiltonian, Eq.~\eqref{H}, can be simulated in an ion trap experiment.
The discussion is a slight generalization of Ref.~\cite{Porras2004a}, to which we refer the reader for details. 
By applying a standing laser wave to a chain of trapped ions, one can arrive (after a canonical transformation) at 
an effective spin model where two internal hyper-fine states act as a pseudo-spin. In this scenario, 
the Coulomb interaction between two ions $i$ and $j$ transmits an effective spin--spin interaction
\begin{equation}
    \label{Jeff}
    J_{ij}^\alpha=-\frac{F_\alpha}{m}\left(\frac{1}{\mathcal{K}^\alpha}\right)_{ij}\,,
\end{equation}
where $F_\alpha$ is the field strength of the laser propagating along direction $\alpha=x,y,z$, $m$ is the ion mass, and the elasticity matrix is
\begin{equation}
    \mathcal{K}_{ij}^{\alpha}=\left\{
                                                        \begin{array}{cc}
                                                            \omega_\alpha^2-c_\alpha \sum_{j'\left(\neq i\right)}\frac{e^2/m}{\left| z_i^0 - z_{j'}^0 \right|^3}\,,&  i=j \\
                                                            +c_\alpha \frac{e^2/m}{\left| z_i^0 - z_{j'}^0 \right|^3}\,, & i\neq j \\
                                                        \end{array}
                                                        \right.
\end{equation}
Here $c_{x,y}=1$, $c_z=-2$, and $\omega_\alpha$ is the vibration frequency of an individual ion.
In the limit where the energy scale given by the Coulomb interaction between neighboring ions becomes much smaller than the energy scale of the vibration of the individual ions $\omega_\alpha$, \emph{i.e.\ }   $\beta_\alpha\equiv\left|c_\alpha\right|e^2/m\omega_\alpha^2 d_0^3\ll 1$, the decay of the spin-spin interactions obeys a dipolar power law. Here, we introduced the mean inter-ion distance $d_0$, and assumed that the ion chain is homogeneous.
In a realistic experimental situation, if an overall trapping potential is used, the inter-ion spacing becomes non-uniform. 
In the central region of a long ion chain, however, the inhomogeneity can be neglected. 
Another way of creating an equidistant chain would consist of placing the ions in individual microtraps \cite{Chiaverini2008}.
For simplicity, we will focus on the homogeneous situation.

Using three pairs of laser waves, one can thus induce the desired dipolar spin interactions in all the three directions of spin space. Choosing the detuning of the laser beams appropriately, one can tune the spin--spin coupling to negative or to positive, as desired.
The missing chemical potential term
can be achieved rather easily via an rf-frequency field driving the transition between the two hyper-fine states which consitute the effective spin $1/2$.
This completes all terms in Hamiltonian~\eqref{H}.

\end{section}


\begin{section}{Expected behavior of the model\label{sec:knownfacts}}

Before proceeding to the detailed numerical analysis of the ground state phase diagram, 
let us briefly sketch the expected behavior of the present model, starting from previous related results. 

For $\theta=0$, Hamiltonian~\eqref{H} 
describes a 1D system of localized hard-core bosons with repulsive dipolar interaction, 
a classical model that can be solved analytically \cite{Bak1982}.
Due to the long-range nature of the interactions, for any given filling factor, the particles arrange in a periodic crystal pattern (a generalized Wigner lattice).
For a given filling factor, these periodic patterns can be constructed by maximizing the mutual distance between the particles. 
Every rational filling factor $q=\frac m n$ occupies a finite extent in $\mu$, thus giving rise to a plateau of fixed particle density, and these plateaux cover the entire range of $\mu$.
Plotting the filling factor against the chemical potential $\mu$ yields a self-similar structure --- a complete devil's staircase, 
similar to Ref.~\cite{Bak1982}. 
The stair's steps have a width
\begin{equation}
        \label{devilsstepwidth}
        \Delta \mu\left(\frac m n\right)=2\frac{3 n \pi^2 \mathrm{csc}\left(\frac{\pi}{n}\right)^2 -n \pi^2 -
        3 \pi^3 \mathrm{cot}\left(\frac{\pi}{n}\right) \mathrm{csc}\left(\frac{\pi}{n}\right)^2}{3 n^3}\,.
\end{equation}
This means that fillings with a large denominator --- equivalent to a large crystal period --- 
become the ground state only in a very small range of chemical potential.

A finite tunneling,  \emph{i.e.}\
$\left(\theta\,\mathrm{mod}\,\pi\right)\neq 0$, introduces quantum mechanics into the system. 
The tunneling allows particles to gain kinetic energy, 
which destabilizes the crystal until it melts at some critical tunneling strength. 
This gives rise to a superfluid (SF) phase where the particles are delocalized over the chain.

There are two possible scenarios for a transition from the crystal structure to the molten phase: 
a direct crystal--SF transition, or an intervening supersolid phase. The latter is an exotic quantum phase 
showing crystal and SF behavior at the same time, which has been predicted for similar --- but two-dimensional --- hard-core boson systems \cite{Wessel2005b,Boninsegni2005b,Heidarian2005,Melko2005,Dang2008,Pollet2010}.
Until recently, there was only one claim of an experimental realization of a supersolid \cite{Kim2004,Kim2004a} in$\phantom{a}^4$He, which is still disputed \cite{Rittner2006,Rittner2007}.
Therefore, it would be interesting to find other systems which show supersolid behavior and are cleaner to interpret. 
One such experiment has been carried out very recently \cite{Baumann2010}, where an atom cloud in a cavity breaks spatial symmetry while at the same time showing off-diagonal long-range order.

In the limit of $\theta=-\pi/2$, Hamiltonian~\eqref{H} describes a ferromagnetic XY model. Here, the most important effects of the long-range nature of the tunneling can be captured in a renormalized nearest-neighbor (NN) interaction: first, all interactions work towards the same ordering, and, second, in one dimension the integral over dipolar interactions converges.
For $\theta=+\pi/2$ the system is a frustrated XY antiferromagnet and the behavior is less obvious. 
It turns out, however, that a similar renormalization to a NN interaction captures the main physics. 
Hence, analogies to a NN-XY model can help understanding the behavior of the system near $\theta=\pm\pi/2$.

In the following, we study how the ground state phase diagram of Hamiltonian~\eqref{H} develops as a function of $\theta$. 
We start with two approximative methods to obtain qualitative insights.

\end{section}


\begin{section}{Mean-field approximations\label{sec:MF}}

\begin{subsection}{Perturbative mean-field theory\label{sec:PMFT}}

A first rough understanding of the crystal--SF transition can be obtained by a perturbative mean-field theory (PMFT), valid for small tunneling. While not being very accurate in 1D, such a mean-field treatment generally becomes better for longer-ranged interactions.

Following the standard recipe of this \emph{Ansatz}, we decompose the Hamiltonian into two parts, $H=H_0+H_1$. 
We choose
\begin{equation}
\label{H0}
H_0=J \sum_{i<j} \frac{1}{\left|i-j\right|^3} S_i^{\,z} S_j^{\,z} -\frac{\mu}{\cos\left(\theta\right)} \sum_i S_i^{\,z}\,,
\end{equation}
\emph{i.e.}~we consider the insulating states, the exact ground states at the devil's staircase ($\theta=0$), as the unperturbed states.
Afterwards, we introduce the perturbation
\begin{eqnarray}
\label{H_1}
    H_1 &=& J \tan\theta \sum_{i<j} \frac{1}{\left|i-j\right|^3} \left( S_i^{\,x} S_j^{\,x} + S_i^{\,y} S_j^{\,y} \right)\\
            &=& J \tan\theta \sum_{i<j} \frac{1}{\left|i-j\right|^3} S_i^{+} S_j^{-} \,. \nonumber
\end{eqnarray}
The aim is now to calculate the expectation value of the SF order parameter
\begin{equation}
\label{varphi}
\varphi_i\equiv\braket{S_i^-}=\mathrm{Tr}\left(\rho S_i^-\right)\,,
\end{equation}
where $S_i^-$ is the spin lowering operator at site $i$ and $\rho$ is the density matrix of the system. 
At the tunneling strength where $\varphi_i$ becomes finite, the assumption that the ground state 
is localized is no longer valid. This gives an upper border for the crystal lobes. 

For the evaluation of the expectation value in Eq.~\eqref{varphi}, we need the density matrix $\rho$, which is given by
\begin{equation}
\label{rho}
\rho=\ue^{-\beta H}/Z \,,
\end{equation}
with $Z$ the partition function and $\beta=1/\left(k_B T\right)$ the inverse temperature. 
We are only interested in ground state properties, equivalent to the limit $\beta\to\infty$. 
In this limit, $Z\to\ue^{-\beta E_0}$, where $E_0$ is the ground state energy.
For small tunneling, \emph{i.e.\ }small $H_1$, 
we can approximate $\rho$ using the Dyson expansion
\begin{equation}
\ue^{-\beta H}\simeq\ue^{-\beta H_0}-\ue^{-\beta H_0} \int_0^\beta \ud \tau \ue^{\tau H_0} H_1 \ue^{-\tau H_0}\,.
\end{equation}
In mean-field approximation, we have
\begin{equation}
\label{H1MF}
H_1\simeq \sum_{i<j} \frac{J \tan\theta}{\left|i-j\right|^3} \left( S_i^{+} \braket{S_j^{-}} + \braket{S_i^{+}} S_j^{-}  -\braket{S_i^{+}}\braket{S_j^{-}} \right)\,.
\end{equation}
Inserting Eqs.~\eqref{rho} to~\eqref{H1MF} into~\eqref{varphi}, we obtain the following self-consistent equations for the SF order parameter
\begin{eqnarray}
\label{varphiselfcons}
\varphi_l &=& -\ue^{\beta E_0} \int_0^\beta\ud\tau \mathrm{Tr} \left\{ S_l^- \ue^{-(\beta-\tau) H_0} \right. \nonumber \\ 
					 & & \qquad \quad \left. \times \left(-\frac{J \tan\theta}{2}\sum_i S_i^+\overline\varphi_i\right)\ue^{-\tau H_0}\right\}\,,
\end{eqnarray}
where we introduced the abbreviation
\begin{equation}
\overline\varphi_i = \sum_{j,j\neq i} \frac{1}{\left|i-j\right|^3} \braket{S_j^{-}}\,.
\end{equation}
It can be checked that in the trace only one-hole excited states and the ground state itself have to be taken into account.
For small $J\tan\theta$, Eq.~\eqref{varphiselfcons} has only the trivial solution $\varphi_i=0$. At some critical $J_c\left(\mu\right)$, however, SF order may develop, \emph{i.e.}\ $\varphi_i\neq 0$. This is equivalent to an instability under adding or removing a single particle. The point where this happens gives an upper border for the crystal phase. The main disadvantage of the current method is that there are potentially more complicated excitations, \emph{e.g.\ }the addition of a particle plus a relocation of the neighboring particles. Such a deformation of the crystal could decrease the potential energy. In the current simple \emph{Ansatz}, however, this type of excitations is not captured.

For our calculations, we assumed an infinite system with the restriction that the states have a periodicity of 12 sites. 
Since the crystal phases with a large periodicity become very small in their extent in $\mu$ [by virtue of Eq.~\eqref{devilsstepwidth}], this is a reasonable restriction which still captures the most prominent features of the phase diagram.

The insulating-lobe structure that we obtain is shown in Fig.~\ref{fig:PMFTlobes}.
The thick line follows the breakdown of the ground state. A dotted line marks the values of $\mu$ where the 
ground state magnetization changes, \emph{i.e.\ }where a different crystal structure becomes lower in energy. 
However, even when a given crystal structure ceases to be the ground state, it can still be a metastable state 
with respect to one-particle or one-hole excitations. In Fig.~\ref{fig:PMFTlobes}, thin lines mark the corresponding 
regions where the most important crystal states are metastable.
For clarity, states that constitute the ground state over only a very small region of the phase diagram have been excluded.
Figure~\ref{fig:PMFTlobes} shows the rich structure of metastable crystal states. 
Recently --- in the context of ultra-cold dipolar neutral atoms --- it has been proposed to use such 
metastable states as quantum memories \cite{Braungardt2007,Pons2007}.
Another feature of the crystal structure is that it is more stable for frustrated antiferromagnetic XY-interaction 
($\theta$>0) than for ferromagnetic XY-interaction ($\theta$<0).
\begin{figure}
\includegraphics[width=0.4\textwidth]{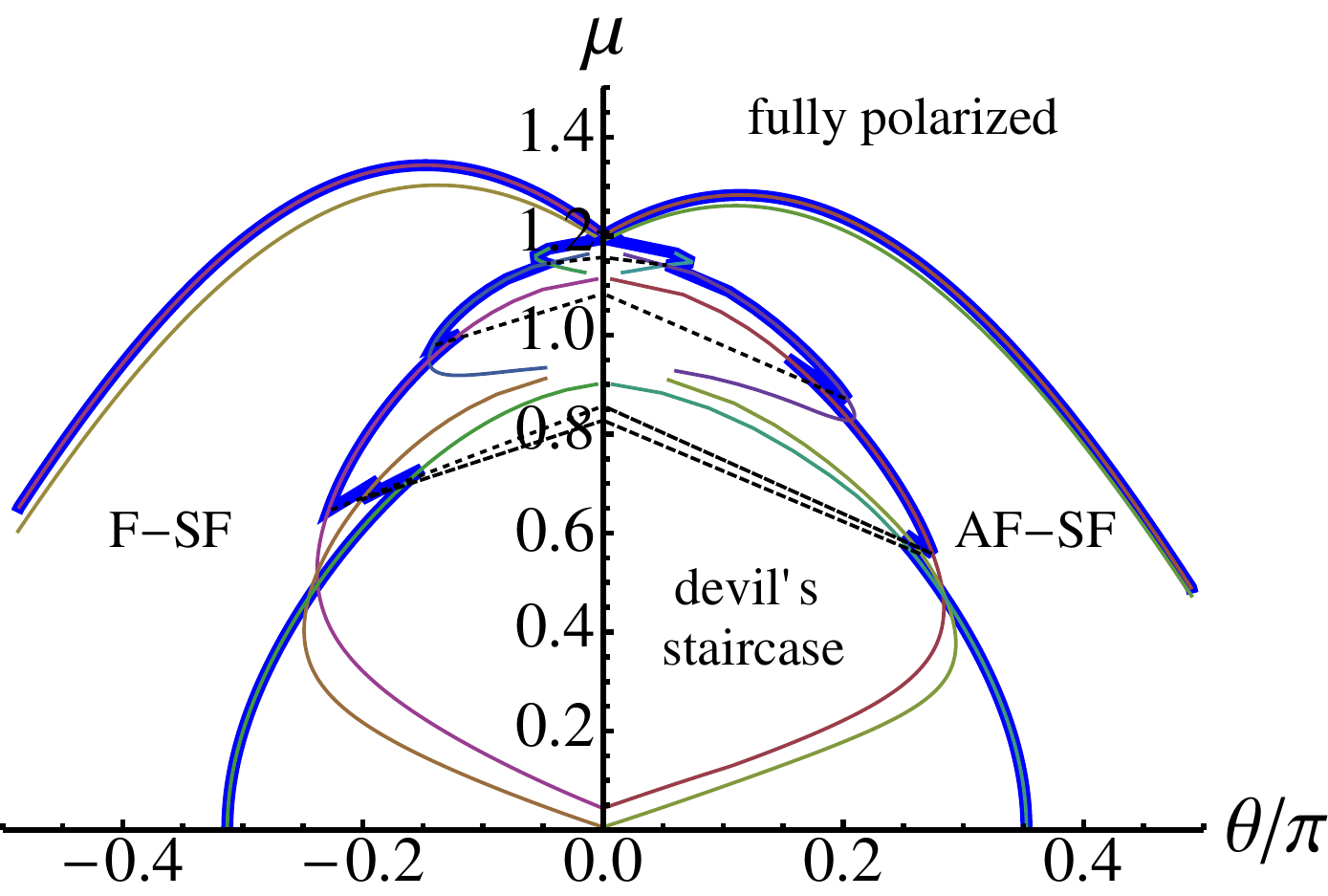}
\caption{
Stability regions of the crystal states within PMFT. The thick line traces the melting of the crystal ground state. The dotted lines mark the values of $\mu$ where the crystal ground state changes periodicity. The thin lines delimit the regions where the crystal states are metastable. Note the asymmetry between the negative (ferromagnetic) and positive (antiferromagnetic) $\theta$ side. F-SF stands for ferromagnetic and AF-SF for antiferromagnetic superfluid.}\label{fig:PMFTlobes}   
\end{figure}

We also computed the phase diagram in a Gutzwiller mean-field theory (not shown). 
The qualitative behavior is similar, but the lobes are somewhat smaller. 
The reason is that PMFT considers destabilization under single-particle or single-hole excitations, 
while Gutzwiller mean-field theory captures better more complicated excitations.

\end{subsection}


\begin{subsection}{Wigner-crystal melting at low filling\label{sec:crystalmelting}}

For very low magnetizations, we can draw an analogy to the melting of Wigner crystals \cite{Arkhipov2005,Buechler2007}.
As described in Sec.~\ref{sec:knownfacts}, at $\theta=0$, the hard-core bosons are perfectly localized, with an inter-particle distance given by the filling fraction. 
At finite tunneling, $\theta\neq0$, the particles spread, but at small tunneling and low filling it is 
reasonable to assume that they remain spatially well separated.
Under this assumption, we can approximate the total wave-function by a product of Gaussians representing the individual particles \footnote{PMFT neglects coherence between different \emph{sites}, while this \emph{Ansatz} neglects coherence between different \emph{particles}.}.
Using this \emph{Ansatz}, one can obtain a self-consistent equation for the distribution of any particle over the lattice.

The spread of the wave packets increases with $\theta$ until the initial assumption that the particles are well separated breaks down, 
and the crystal has to be considered as molten.
A measure for this is the normalized variance 
$L\equiv\frac{\Delta x}{d_0}=\frac{1}{d_0}\sqrt{\braket{x^2}-\braket{x}^2}$, called the Lindemann parameter \cite{Lindemann1910}.
As shown in Fig.~\ref{fig:CMLindemann},
\begin{figure}
 \includegraphics[width=0.35\textwidth]{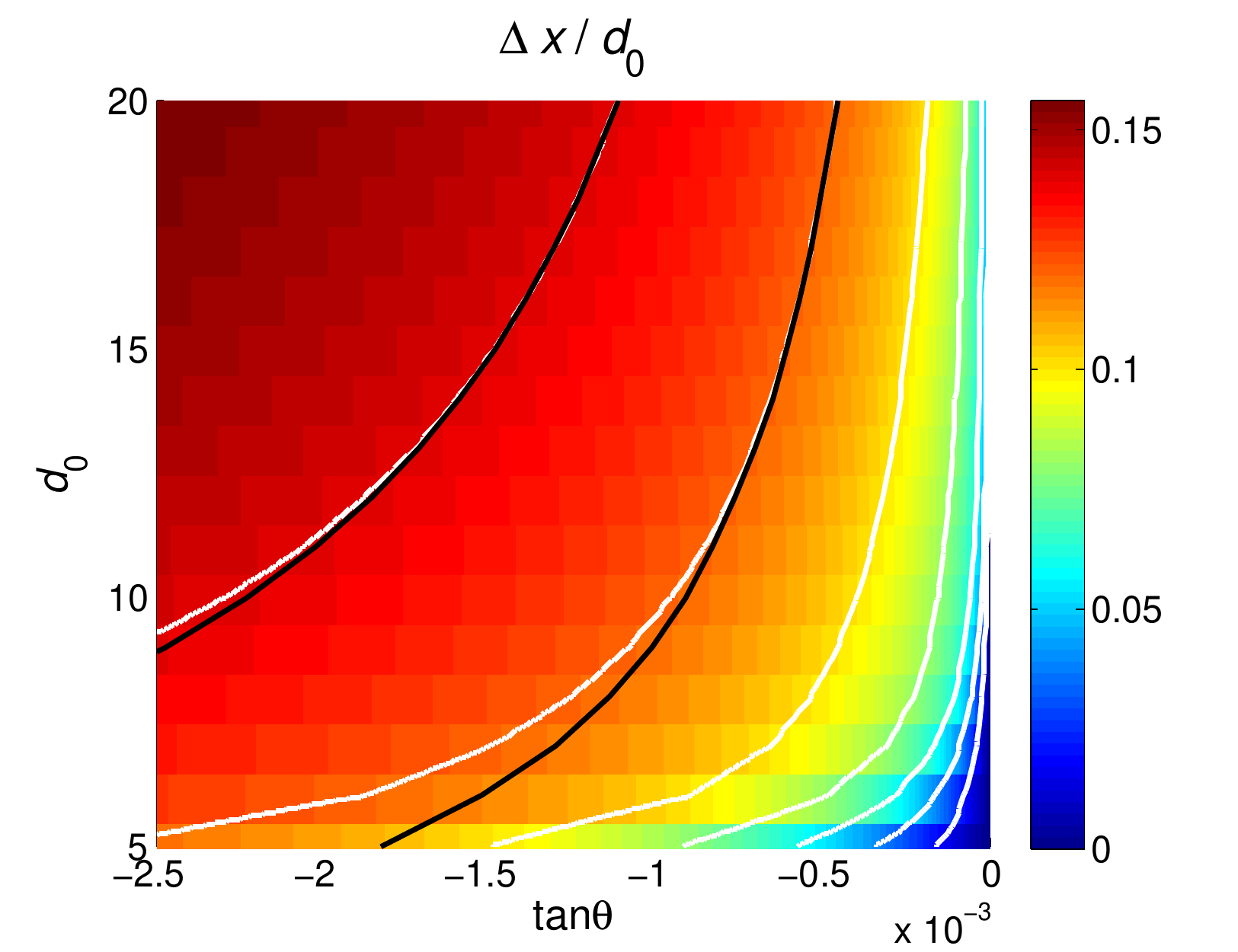}
 \caption{
 \label{fig:CMLindemann}
 The Lindemann parameter (described in the text) as a function of the interaction parameter $\theta$ and 
 the mean distance between particles in the crystalline phase. The white lines are contour lines at $L=0.02\dots 0.14$ in steps of $0.02$. For comparison two black lines following the law $\tan\theta\propto d_0^{-1}$ are also shown.
 }
\end{figure}
for large inter-particle spacing $d_0$, the system behaves similar to the continuum limit, where it is known \cite{Arkhipov2005} that there is a melting transition at $J \tan\theta_c=\mathrm{const}/d_0$.
We find that the contour lines of the Lindemann parameter from our crude approximation follow this behavior well for large $d_0$. 
At small inter-particle distances there are deviations, but here the assumptions made in our approximation are not valid anyway.
The physical explanation behind the observed behavior is that for large inter-particle distances the repulsive interaction becomes very weak. 
Hence, for larger $d_0$ a smaller gain in kinetic energy suffices to destabilize the crystal.

\end{subsection}

\end{section}


\begin{section}{Density-matrix renormalization group (DMRG)\label{sec:DMRG}}

While mean-field theories can provide some physical understanding of the system, in 1D they are not very precise. 
Therefore, we turn now to a thorough numerical 
analysis of the phase diagram by the quasi-exact density-matrix renormalization group (DMRG) method \cite{alps,White1992b,Schollwock2005}.
We consider chains with open boundary conditions of up to $N=102$ sites, and with a bond dimension of $D=128$.
The interaction range was always cut-off at half the length of the chain. 

\begin{figure}
\includegraphics[width=0.5\textwidth]{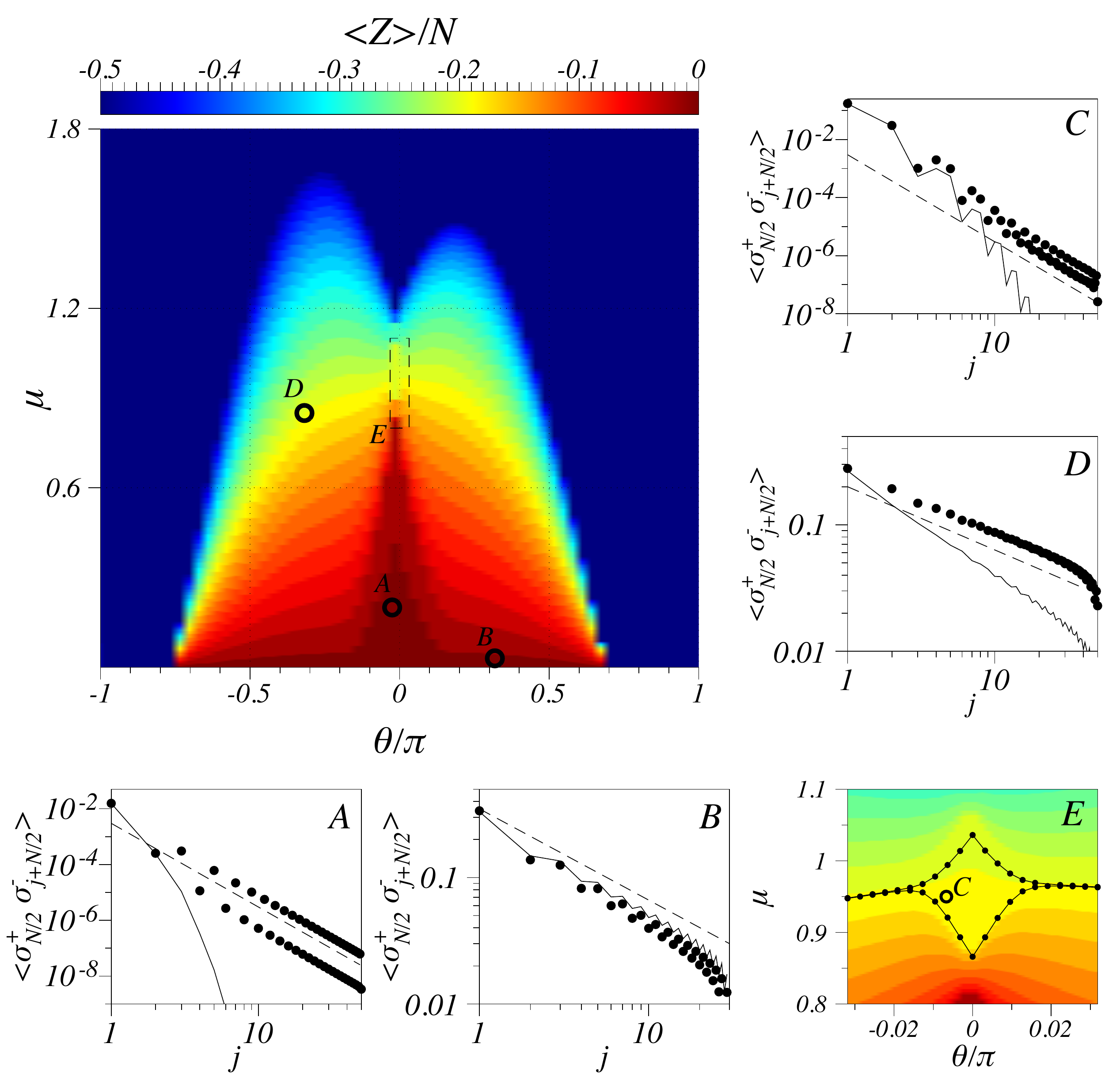}
\caption{
					\label{fig:DMRGPhaseDiagram}
					Mean polarization $\braket{Z}/N\equiv\frac{1}{2}\sum_{j=1}^N\braket{\sigma_j^z}/N$ for a system with dipolar hopping and $N=60$. 
					The log--log plots in Panels A to D show, for $N=102$, the decay of the in-plane correlations $G\left(j\right)$ at the marked points of the phase diagram, comparing dipolar hopping (points) with NN hopping (solid lines). The dashed lines are guides to eye proportional to $j^{-3}$.
					For dipolar hopping, the decay is algebraic everywhere, even within the lobes (A and C), where it follows the dipolar exponent $\alpha=3$, while for NN hopping the decay is exponential. 
					In the SF (B and D), the algebraic decay is faster than for NN hopping on the antiferromagnetic side and slower on the ferromagnetic side. 
					Panel E shows a zoom on the dashed section together with the finite-size scaling result of the crystal lobe at $1/3$ filling, as explained in the text.  			
					}
					
\end{figure}

Figure~\ref{fig:DMRGPhaseDiagram} presents the mean polarization 
$\braket{Z}/N\equiv\frac{1}{2}\sum_{j=1}^N\braket{\sigma_j^z}/N$, 
where $N$ is the number of sites and $\sigma_j^z$ the Pauli-$z$ matrix of spin $j$. 
We find that for large enough field $\mu$, or strong ferromagnetic ZZ-interaction 
(corresponding to $\theta$ not too far from $\pm \pi$), 
the system is in a fully polarized state, quantitatively similar to what we saw in the mean-field calculations.
However, as expected, the general precision of the mean-field calculations in this one-dimensional system is low: 
the DMRG results indicate that the $\theta$-range of the crystal lobes is up to an order of magnitude smaller than predicted in mean-field. 
In fact, in the global view given in the main panel of Fig.~\ref{fig:DMRGPhaseDiagram}, only the plateau for $1/2$ filling is discernible.
Panel E of Fig.~\ref{fig:DMRGPhaseDiagram} 
shows the small region of the phase diagram where the $1/3$-filling crystal lobe is located.
Due to the long-range nature of the hopping terms, the polarization is asymmetric with respect to $\pm\theta$, in contrast to \cite{Burnell2009b}.

We find that open boundary conditions play an important role for the finite systems used here. 
For example, in the center of the main panel of Fig.~\ref{fig:DMRGPhaseDiagram}, 
there appears a broad plateau with $(1/2-1/N)$ filling (\emph{i.e.\ }with one spin flipped away from half filling). 
In the thermodynamic limit, this phase is not as stable and merges with the $1/2$-filling plateau to a single lobe. 
This peculiarity is highly relevant for the correct interpretation of future experiments, 
which might be carried out with open boundary conditions for technical convenience. 
Moreover, boundary effects play an important role in the determination of the decay of correlations, which we now turn to describe. 

The log--log plots in panels A to D show the decay of the in-plane correlations 
$G\left(j\right)\equiv\left|\braket{\sigma_{N/2}^+\sigma_{N/2+j}^-}\right|$ 
at some selected points in the phase diagram (filled circles), and compare the result to the same system but
with nearest-neighbor (NN) hopping (solid line).
In the superfluid phase (panels B and D), the decay is algebraic for both NN and dipolar hopping. 
Panels A and C lie inside the lobes corresponding to $1/2$ and $1/3$ filling, respectively, which is visible through the oscillation of the correlation function superposed with an overall decay. 
We see that, for dipolar hopping, the decay is algebraic everywhere, even within the lobes, 
where for NN hopping the correlations decay exponentially. In the lobes, the decay for dipolar hopping 
follows the interactions with exponent $\alpha=3$ \cite{Deng2005}. Notice that a clear fit for this exponent can only be 
obtained for rather large chains with $N>60$ spins. 
Even more, although the strength of the XY-interactions decays very rapidly with distance, the power-law decay of correlations inside the lobes 
cannot be observed at all if the interaction is truncated.

Since inside the lobes we have a crystal structure, the long-range decay of the transverse correlations 
is an anomaly when compared to other models with insulating phases. 
For this similarity to a supersolid, we call this phase a quasi-supersolid phase. In a true supersolid, a crystal structure coexists with a long-range order 
in the transverse correlations. However, in 1D off-diagonal LRO cannot be spontaneously broken \cite{Mermin1966}. Hence, the closest analogue to
a supersolid that can exist in 1D is a state as the present one --- with diagonal LRO and algebraically decaying off-diagonal correlations. 

\begin{figure}
\includegraphics[width=0.5\textwidth]{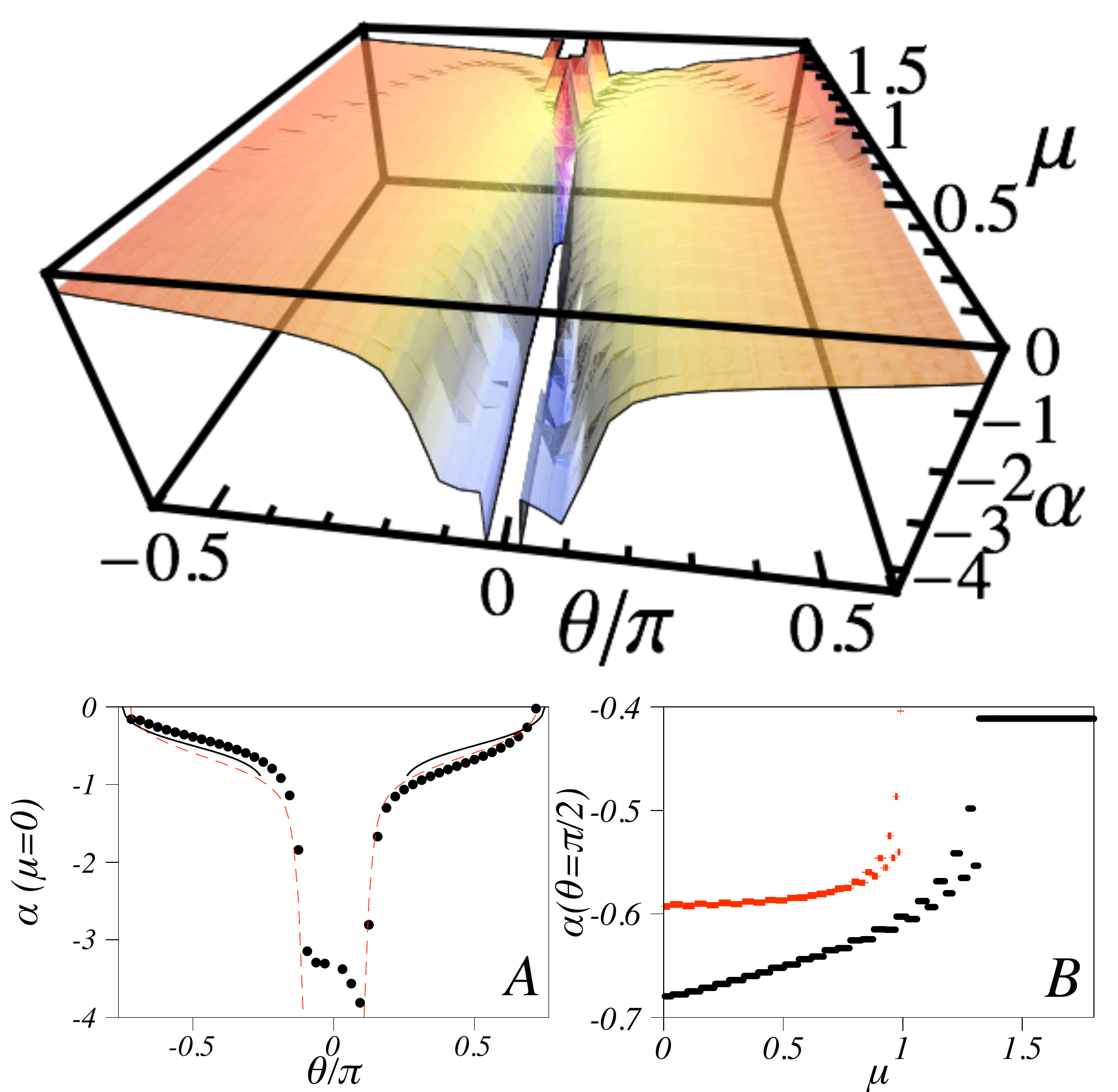}
\caption{
					Exponents $\alpha$ of an algebraic fit $G_a\left(j\right)=c_0 j^\alpha$ to the correlations $G\left(j\right)$, for $N=60$. 
					The panels show cuts at $\mu=0$ ($1/2$ filling, A) and $\theta=\pi/2$ (XY model, B).
					The black lines in A are the exact results for the NN XXZ model. 
					Antiferromagnetic interaction induces larger exponents compared to the NN XXZ model, while ferromagnetic interaction leads to smaller exponents. 						Inside the insulating lobes, $\alpha=3$ (deviations are due to finite size effects).
					Results for NN tunneling are shown in red (dashed line in A, filled circles in B).
						}\label{fig:DMRGExponents}
\end{figure}

A change in the behavior of the correlations marks the transition from a crystal state to the SF. 
For a quantitative evaluation, at each point in the phase diagram we fit an algebraic trial function $G_a\left(j\right)=c_0 j^{-\alpha}$.
The value of $\alpha$ for a chain of $N=60$ spins is shown in Fig.~\ref{fig:DMRGExponents}.
The cuts of Fig.~\ref{fig:DMRGExponents} at $\mu=0$ ($1/2$ filling, A) and $\theta=\pi/2$ (XY model, B) 
show the wide range that the exponent of the correlations takes in this system. 
They also demonstrate the influence of frustration: the correlations decay faster at the antiferromagnetic side ($\theta>0$) than at the ferromagnetic side ($\theta<0$).
The exponents are similar to the NN XXZ model (black line in A), but cover a broader range. In particular, they are not independent of the filling for $\theta=\pi/2$ (panel B), where the NN XXZ model has $\alpha=-0.5$. 
The results for a system with dipolar ZZ-interaction but NN XY-interaction are shown in red (dashed line in A, filled circles in B). The main qualitative difference is that inside the insulating lobes the decay is no longer a power law, but follows an exponential. 

In a finite system, finding the phase transition from the crystal state to the SF is not simple,
because the finite number of spins prevents the system from assuming arbitrary polarizations. This results in a division of the phase diagram into different stripes with fixed integer number of up spins, which makes the crystal lobes difficult to discern. 
In infinite systems, however, one expects a step-like behavior of the polarization 
only in crystal phases, while in the SF the polarization
should change smoothly with $\mu$ between $-1/2$ and $1/2$.
This observation makes it possible to extrapolate the border between the crystal phases and the SF by the following finite size scaling: 
At fixed $\theta$ and for a given polarization, 
compute for several chain lengths $N$ the upper and lower limits of the polarization plateau, 
$\mu_a\left(N\right)$ and $\mu_b\left(N\right)$. 
Then, from a finite size scaling of the results, one can extract $\mu_{a,b}\left(N=\infty\right)$. 
In the SF, where the polarization should be a continuum, $\mu_{a}\left(\infty\right)$ and $\mu_{b}\left(\infty\right)$
should be equal. However, in the crystal lobe there will be a finite distance between $\mu_a\left(\infty\right)$ and $\mu_b\left(\infty\right)$:
the width of the lobe for that value of $\theta$.
A similar procedure is known to work well for the estimation of Mott-lobes from finite systems in the Bose--Hubbard model 
\cite{Andreprivate}. The main difference is that in our model there is in principle a lobe for any rational filling factor, instead of 
only for integer filling factors. 
Panel~E of Fig.~\ref{fig:DMRGPhaseDiagram} shows the result of this approach for the $1/3$-filling lobe, using chain lengths of up to 102 spins. 
The cusp structure is typical for one dimensional systems. 

\end{section}


\begin{section}{ Infinite time evolving block decimation (iTEBD)\label{sec:iTEBD}}

  To complement the results of the previous section, we study
  the phase diagram with the iTEBD algorithm~\cite{Vidal2007},
  in which infinite chains can be directly addressed without the need
  of finite size extrapolations. Treating long-range interaction
  complicates the original formulation of the iTEBD algorithm and
  turns out to lead to convergence problems. Instead, we implement a
  variant of the original algorithm in which the translationally
  invariant character of the problem is kept in the state and the
  interactions, thanks to the use of Matrix Product
  Operators~\cite{Murg2008a}.  In this way, we can include interaction
  terms ranging longer than nearest neighbors in a much simpler way.
  
  By comparing the phase diagrams for different range of XY
  interactions, we may study the effect of the long-range hopping in
  the thermodynamic limit.  The method still requires a truncation of
  the interacion range to some finite order, so that it will not be
  possible to reproduce the power-law decay of correlations within the
  crystal lobes.

  We truncate the ZZ interactions above next-to-nearest-neighbor (NNN) interactions and compare the cases
  of NN and up to NNN XX and YY terms.
  We observe that a small bond dimension, $D=10$, provides already a
good approximation to the overall phase diagram. To analyze the isolating lobe at $1/3$ filling
we increase the bond dimension to $D=20$.

  Figure~\ref{fig:pd112} shows the results for the magnetization per
  particle in the first case, in which XX and YY interactions range
  only to the NN.  As in~\cite{Burnell2009b} the phase
  diagram is symmetric. This is also seen in the close up on
  the filling $1/3$ crystal lobe (right panel of
  Fig.~\ref{fig:pd112}).

  Figure~\ref{fig:pd222} shows the equivalent phase diagram when
  NNN terms are included in the XY plane.
  Including one more term in the XY interactions already causes a
  clear asymmetry of the phase diagram with respect to the change
  $\theta \rightarrow -\theta$.  This is clearly visible in the
  superfluid phases (left panel of the figure), and the zoom on the
  isolating lobe at polarization $-1/6$ shows some asymmetry as well.
  
  The size of the lobe is larger than the result from DMRG after
  finite size extrapolation. We are, however, only including the first
  term further than the NN, and we expect that longer
  range terms correct the exact shape of the lobe, as they will give rise to
  lobes corresponding to other filling factors.

  \begin{figure}
    \includegraphics[width=0.5\textwidth]{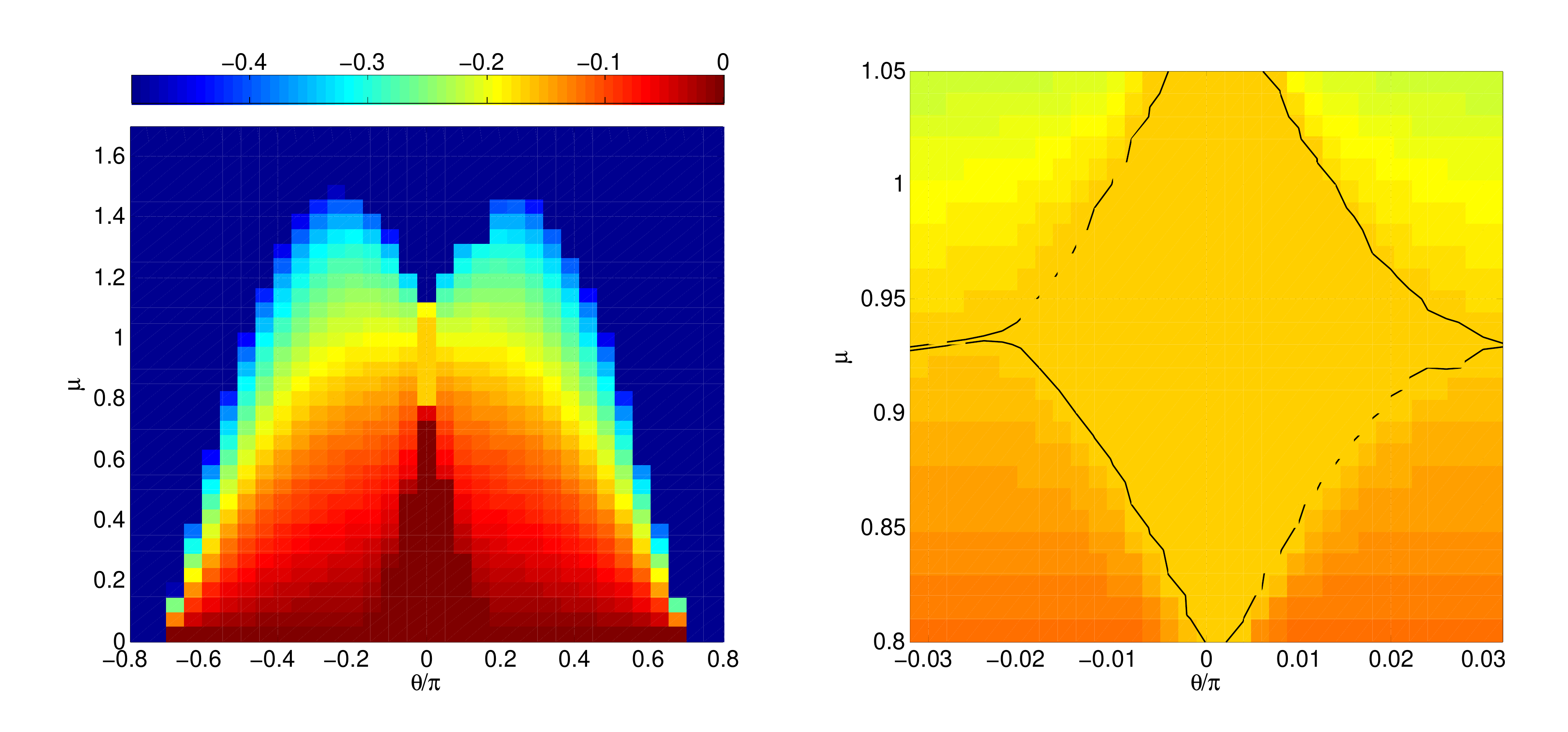}
    \caption{Mean magnetization for an infinite chain with
      NN XY interactions and NNN ZZ
      term. The right panel shows a zoom on the isolating lobe at
      $1/3$ filling.  }
    \label{fig:pd112}
  \end{figure}

  \begin{figure}
    \includegraphics[width=0.5\textwidth]{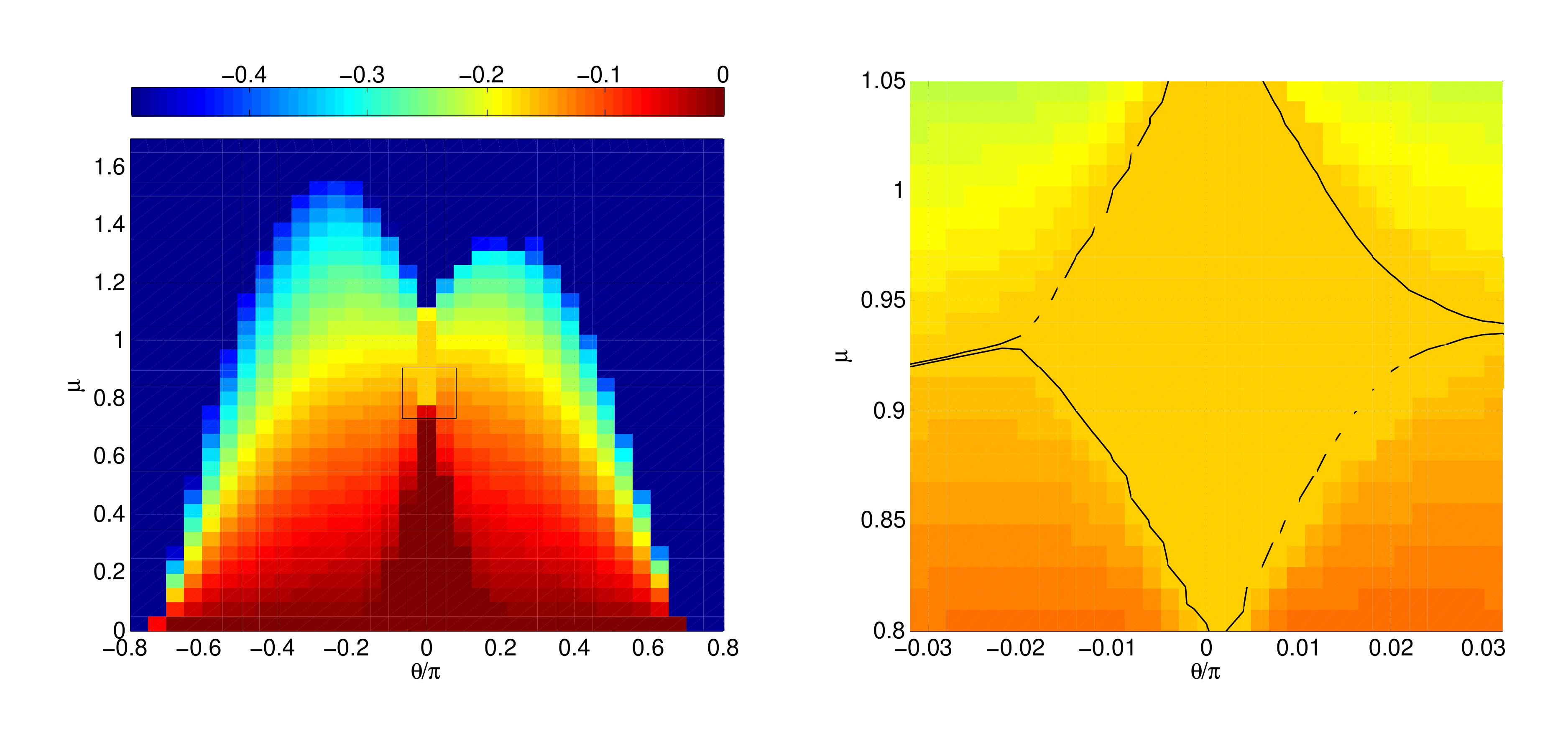}
    \caption{ Mean magnetization for an infinite chain when the range
      of all ZZ and XY interactions is extended to
      the NNN. The right panel shows a zoom on the
      isolating lobe at $1/3$ filling.  }
    \label{fig:pd222}
  \end{figure}

\end{section}


\begin{section}{Exact diagonalization\label{sec:ED}}

In this section, we supplement the previous results by exact diagonalization (ED) of chains of $18$ spins with periodic boundary conditions, using the Lanczos method. Although ED only allows investigation of very small systems, from these one can often infer surprisingly much about the behavior of the model at larger sizes.
Moreover, an experimental implementation with a chain of trapped ions will have to start with short chain lengths. For such a case ED provides a valuable validation of the trapped-ion quantum simulator.

Figure~\ref{fig:plotfid} shows the fidelity susceptibility, 
\begin{equation}
\chi_F=\frac{1-\left|\braket{\psi\left(\theta\right)|\psi\left(\theta+\Delta\theta\right)}\right|^2}{\Delta\theta^2}
\end{equation} 
for $1/3$ filling. At a second order quantum phase transition, $\chi_F$ diverges in the thermodynamic limit. 
The peaks in a finite system are precursors thereof. However, in the present system the divergence 
turns out to be relatively weak, which makes it difficult to extrapolate quantitative values for 
the critical point of the transition. The qualitative value, however, is consistent with the DMRG and iTEBD analyses. 
The second derivative of the energy has been proposed \cite{Chen2008} 
as a substitute of the fidelity susceptibility to detect quantum phase transitions.
In the present system, however, the energy is very smooth and does not give a good indication of the phase transition.

Instead, we compare in Fig.~\ref{fig:plotfid} $\chi_F$ to the second derivative of the total in-plane magnetization, 
$\partial^2 M_{x}/\partial\theta^2$, where we defined $M_{x}\equiv\sum_{i\neq j}\left|\braket{S_i^x S_j^x + S_i^y S_j^y}\right|/\left(N\left(N-1\right)/2\right)$. Here, the prime indicates that the sum excludes self-correlations.
We take the absolute value of the correlations in order to take into account the different staggered 
configurations occurring for different filling factors.
We see that $\partial^2 M_{x}/\partial\theta^2$ peaks at slightly larger absolute values of $\theta$ than $\chi_F$, but the rough locations are consistent with the peaks of $\chi_F$ \footnote{We find that the second derivative of the total out-of-plane magnetization, $M_z\equiv\sum_{i\neq j}\left|\braket{S_i^z S_j^z}\right|/\left(N\left(N-1\right)/2\right)$, (not shown) behaves very similar to $\partial^2 M_{x}/\partial\theta^2$.}.

A comparison to the results from DMRG shows a very similar qualitative behavior. 
This means that already the very small chains considered here show strong precursors of the physics relevant for large chain lengths.

\begin{figure}
\includegraphics[width=0.5\textwidth]{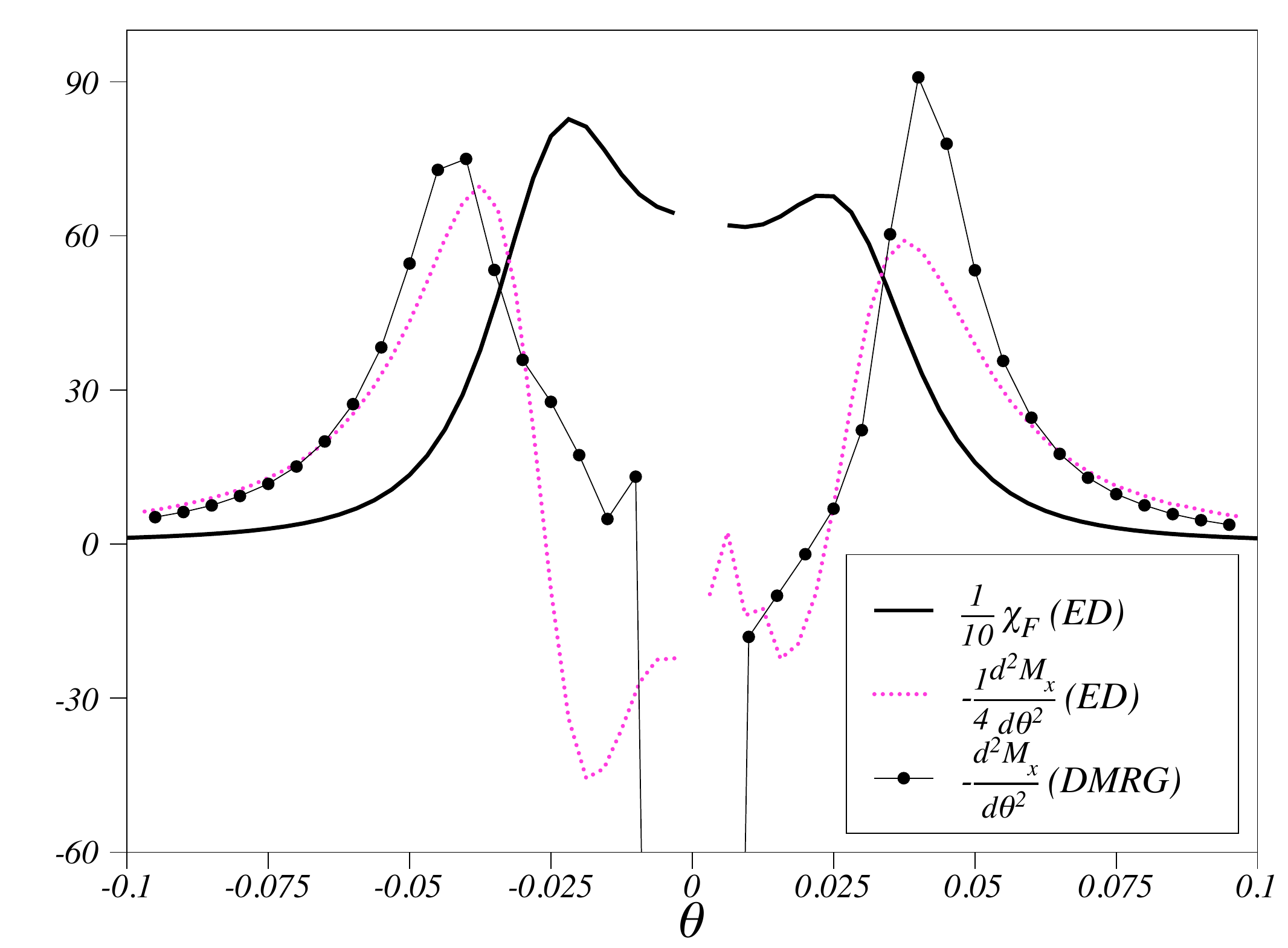}
\caption{
Fidelity susceptibility $\chi_F/10$ for 18 spins (ED), and second derivative of total in-plane magnetization $\partial^2 M_{x}/\partial\theta^2$ for 18 spins (ED) and for 60 spins (DMRG). 
\label{fig:plotfid}
}
\end{figure}

\end{section}


\begin{section}{Conclusion\label{sec:conclusion}}

To summarize, we have studied a one dimensional system of hard-core bosons where particles interact and can tunnel at long 
distances with an algebraically decaying strength. We have discussed how the system can be mapped onto
a spin Hamiltonian that can be simulated experimentally using trapped ions. Unlike other atomic quantum simulators
(such as dipolar ultracold atoms), trapped ions appear to be more flexible in the manipulation of
some parameters --- \emph{e.g.}~it is easy to change the interactions and hoppings from negative to positive
simply by detuning the lasers.

The system we have studied has a rich phase diagram, with many insulating phases filled with (possibly
long-lived) meta-stable states and quasi-long range order in all correlations --- two prominent features
that are induced solely by the long-range nature of the interactions. 

The coexistence of diagonal LRO and off-diagonal quasi-LRO in the insulating lobes can be seen as a quasi-supersolid, the closest analogue to a supersolid that can be found in 1D. 
It would be very interesting to study how these phases behave in 2D, where the possibility of having off-diagonal LRO could allow the quasi-supersolids to become true supersolids.

Since calculations prove to be quite complicated for long-range interactions, one can consider the present model as a testbed for a comparison of different numerical tools.
We find that the precision of the standard perturbative mean-field theory is insufficient, because it does not capture excitations more complicated than single particle excitations. Exact diagonalization, DMRG, and iTEBD, however, all deliver a consistent picture.
Among these, we find that DMRG is the method of choice for the model studied. However, it is restricted to finite systems. While iTEBD is less accurate and restricted to shorter range of interactions, it allows to investigate the thermodynamic limit directly. Exact diagonalization, finally, while being relevant only for very small systems, can still serve for benchmarking other less accurate computational methods and near-future experiments. 

In the future, it would be interesting to pursue investigations of other properties of the system that are affected by the long-range nature of the interactions, such as its response to excitations or the dynamics of its correlations. 

\end{section}

\begin{section}{Acknowledgments}
We thank Andr\'{e} Eckardt and Christian Trefzger for stimulating discussions. 
This work is financially supported by the Caixa Manresa, Spanish MICINN (FIS2008-00784 (TOQATA) and Consolider QOIT), EU Integrated Project AQUTE, ERC Advanced Grant QUAGATUA, and the EU STREP NAMEQUAM.
\end{section}


\bibliographystyle{plain}

\begin{thebibliography}{40}
\expandafter\ifx\csname natexlab\endcsname\relax\def\natexlab#1{#1}\fi
\expandafter\ifx\csname bibnamefont\endcsname\relax
  \def\bibnamefont#1{#1}\fi
\expandafter\ifx\csname bibfnamefont\endcsname\relax
  \def\bibfnamefont#1{#1}\fi
\expandafter\ifx\csname citenamefont\endcsname\relax
  \def\citenamefont#1{#1}\fi
\expandafter\ifx\csname url\endcsname\relax
  \def\url#1{\texttt{#1}}\fi
\expandafter\ifx\csname urlprefix\endcsname\relax\def\urlprefix{URL }\fi
\providecommand{\bibinfo}[2]{#2}
\providecommand{\eprint}[2][]{\url{#2}}

\bibitem[{\citenamefont{Lahaye et~al.}(2009)\citenamefont{Lahaye, Menotti,
  Santos, Lewenstein, and Pfau}}]{Lahaye2009}
\bibinfo{author}{\bibfnamefont{T.}~\bibnamefont{Lahaye}},
  \bibinfo{author}{\bibfnamefont{C.}~\bibnamefont{Menotti}},
  \bibinfo{author}{\bibfnamefont{L.}~\bibnamefont{Santos}},
  \bibinfo{author}{\bibfnamefont{M.}~\bibnamefont{Lewenstein}},
  \bibnamefont{and} \bibinfo{author}{\bibfnamefont{T.}~\bibnamefont{Pfau}},
  \bibinfo{journal}{Rep. Prog. Phys.} \textbf{\bibinfo{volume}{72}},
  \bibinfo{pages}{126401} (\bibinfo{year}{2009}).

\bibitem[{\citenamefont{Capogrosso-Sansone
  et~al.}(2010)\citenamefont{Capogrosso-Sansone, Trefzger, Lewenstein, Zoller,
  and Pupillo}}]{Capogrosso2010}
\bibinfo{author}{\bibfnamefont{B.}~\bibnamefont{Capogrosso-Sansone}},
  \bibinfo{author}{\bibfnamefont{C.}~\bibnamefont{Trefzger}},
  \bibinfo{author}{\bibfnamefont{M.}~\bibnamefont{Lewenstein}},
  \bibinfo{author}{\bibfnamefont{P.}~\bibnamefont{Zoller}}, \bibnamefont{and}
  \bibinfo{author}{\bibfnamefont{G.}~\bibnamefont{Pupillo}},
  \bibinfo{journal}{Phys. Rev. Lett.} \textbf{\bibinfo{volume}{104}},
  \bibinfo{pages}{125301} (\bibinfo{year}{2010}).

\bibitem[{\citenamefont{Batrouni et~al.}(1995)\citenamefont{Batrouni,
  Scalettar, Zimanyi, and Kampf}}]{Batrouni1995}
\bibinfo{author}{\bibfnamefont{G.~G.} \bibnamefont{Batrouni}},
  \bibinfo{author}{\bibfnamefont{R.~T.} \bibnamefont{Scalettar}},
  \bibinfo{author}{\bibfnamefont{G.~T.} \bibnamefont{Zimanyi}},
  \bibnamefont{and} \bibinfo{author}{\bibfnamefont{A.~P.} \bibnamefont{Kampf}},
  \bibinfo{journal}{Phys. Rev. Lett.} \textbf{\bibinfo{volume}{74}},
  \bibinfo{pages}{2527} (\bibinfo{year}{1995}).

\bibitem[{\citenamefont{Sengupta et~al.}(2005)\citenamefont{Sengupta, Pryadko,
  Alet, Troyer, , and Schmid}}]{Sengupta2005}
\bibinfo{author}{\bibfnamefont{P.}~\bibnamefont{Sengupta}},
  \bibinfo{author}{\bibfnamefont{L.~P.} \bibnamefont{Pryadko}},
  \bibinfo{author}{\bibfnamefont{F.}~\bibnamefont{Alet}},
  \bibinfo{author}{\bibfnamefont{M.}~\bibnamefont{Troyer}}, , \bibnamefont{and}
  \bibinfo{author}{\bibfnamefont{G.}~\bibnamefont{Schmid}},
  \bibinfo{journal}{Phys. Rev. Lett.} \textbf{\bibinfo{volume}{94}},
  \bibinfo{pages}{207202} (\bibinfo{year}{2005}).

\bibitem[{\citenamefont{Dang et~al.}(2008)\citenamefont{Dang, Boninsegni, and
  Pollet}}]{Dang2008}
\bibinfo{author}{\bibfnamefont{L.}~\bibnamefont{Dang}},
  \bibinfo{author}{\bibfnamefont{M.}~\bibnamefont{Boninsegni}},
  \bibnamefont{and} \bibinfo{author}{\bibfnamefont{L.}~\bibnamefont{Pollet}},
  \bibinfo{journal}{Phys. Rev. B} \textbf{\bibinfo{volume}{78}},
  \bibinfo{pages}{132512} (\bibinfo{year}{2008}).

\bibitem[{\citenamefont{Menotti et~al.}(2007)\citenamefont{Menotti, Trefzger,
  and Lewenstein}}]{Menotti2007b}
\bibinfo{author}{\bibfnamefont{C.}~\bibnamefont{Menotti}},
  \bibinfo{author}{\bibfnamefont{C.}~\bibnamefont{Trefzger}}, \bibnamefont{and}
  \bibinfo{author}{\bibfnamefont{M.}~\bibnamefont{Lewenstein}},
  \bibinfo{journal}{Phys. Rev. Lett.} \textbf{\bibinfo{volume}{98}},
  \bibinfo{pages}{235301} (\bibinfo{year}{2007}).

\bibitem[{\citenamefont{Trefzger et~al.}(2008)\citenamefont{Trefzger, Menotti,
  and Lewenstein}}]{Trefzger2008}
\bibinfo{author}{\bibfnamefont{C.}~\bibnamefont{Trefzger}},
  \bibinfo{author}{\bibfnamefont{C.}~\bibnamefont{Menotti}}, \bibnamefont{and}
  \bibinfo{author}{\bibfnamefont{M.}~\bibnamefont{Lewenstein}},
  \bibinfo{journal}{Phys. Rev. A} \textbf{\bibinfo{volume}{78}},
  \bibinfo{eid}{043604} (\bibinfo{year}{2008}).

\bibitem[{\citenamefont{Braungardt et~al.}(2007)\citenamefont{Braungardt,
  Sen(De), Sen, and Lewenstein}}]{Braungardt2007}
\bibinfo{author}{\bibfnamefont{S.}~\bibnamefont{Braungardt}},
  \bibinfo{author}{\bibfnamefont{A.}~\bibnamefont{Sen(De)}},
  \bibinfo{author}{\bibfnamefont{U.}~\bibnamefont{Sen}}, \bibnamefont{and}
  \bibinfo{author}{\bibfnamefont{M.}~\bibnamefont{Lewenstein}},
  \bibinfo{journal}{Phys. Rev. A} \textbf{\bibinfo{volume}{76}},
  \bibinfo{eid}{042307} (\bibinfo{year}{2007}).

\bibitem[{\citenamefont{Pons et~al.}(2007)\citenamefont{Pons, Ahufinger,
  Wunderlich, Sanpera, Braungardt, Sen(De), Sen, and Lewenstein}}]{Pons2007}
\bibinfo{author}{\bibfnamefont{M.}~\bibnamefont{Pons}},
  \bibinfo{author}{\bibfnamefont{V.}~\bibnamefont{Ahufinger}},
  \bibinfo{author}{\bibfnamefont{C.}~\bibnamefont{Wunderlich}},
  \bibinfo{author}{\bibfnamefont{A.}~\bibnamefont{Sanpera}},
  \bibinfo{author}{\bibfnamefont{S.}~\bibnamefont{Braungardt}},
  \bibinfo{author}{\bibfnamefont{A.}~\bibnamefont{Sen(De)}},
  \bibinfo{author}{\bibfnamefont{U.}~\bibnamefont{Sen}}, \bibnamefont{and}
  \bibinfo{author}{\bibfnamefont{M.}~\bibnamefont{Lewenstein}},
  \bibinfo{journal}{Phys. Rev. Lett.} \textbf{\bibinfo{volume}{98}},
  \bibinfo{eid}{023003} (\bibinfo{year}{2007}).

\bibitem[{\citenamefont{Mukamel}(2009)}]{Mukamel2009}
\bibinfo{author}{\bibfnamefont{D.}~\bibnamefont{Mukamel}},
  \bibinfo{journal}{arXiv:0905.1457v1}  (\bibinfo{year}{2009}).

\bibitem[{\citenamefont{Porras and Cirac}(2004)}]{Porras2004a}
\bibinfo{author}{\bibfnamefont{D.}~\bibnamefont{Porras}} \bibnamefont{and}
  \bibinfo{author}{\bibfnamefont{J.~I.} \bibnamefont{Cirac}},
  \bibinfo{journal}{Phys. Rev. Lett.} \textbf{\bibinfo{volume}{92}},
  \bibinfo{pages}{207901} (\bibinfo{year}{2004}).

\bibitem[{\citenamefont{Mintert and Wunderlich}(2001)}]{Mintert2001}
\bibinfo{author}{\bibfnamefont{F.}~\bibnamefont{Mintert}} \bibnamefont{and}
  \bibinfo{author}{\bibfnamefont{C.}~\bibnamefont{Wunderlich}},
  \bibinfo{journal}{Phys. Rev. Lett.} \textbf{\bibinfo{volume}{87}},
  \bibinfo{pages}{257904} (\bibinfo{year}{2001}).

\bibitem[{\citenamefont{Friedenauer et~al.}(2008)\citenamefont{Friedenauer,
  Schmitz, Glueckert, Porras, and Schaetz}}]{Friedenauer2008}
\bibinfo{author}{\bibfnamefont{A.}~\bibnamefont{Friedenauer}},
  \bibinfo{author}{\bibfnamefont{H.}~\bibnamefont{Schmitz}},
  \bibinfo{author}{\bibfnamefont{J.~T.} \bibnamefont{Glueckert}},
  \bibinfo{author}{\bibfnamefont{D.}~\bibnamefont{Porras}}, \bibnamefont{and}
  \bibinfo{author}{\bibfnamefont{T.}~\bibnamefont{Schaetz}},
  \bibinfo{journal}{Nat. Phys.} \textbf{\bibinfo{volume}{4}},
  \bibinfo{pages}{757} (\bibinfo{year}{2008}).

\bibitem[{\citenamefont{Kim et~al.}(2010)\citenamefont{Kim, Chang, Korenblit,
  Islam, Edwards, Freericks, Lin, Duan, and Monroe}}]{Kim2010}
\bibinfo{author}{\bibfnamefont{K.}~\bibnamefont{Kim}},
  \bibinfo{author}{\bibfnamefont{M.-S.} \bibnamefont{Chang}},
  \bibinfo{author}{\bibfnamefont{S.}~\bibnamefont{Korenblit}},
  \bibinfo{author}{\bibfnamefont{R.}~\bibnamefont{Islam}},
  \bibinfo{author}{\bibfnamefont{E.~E.} \bibnamefont{Edwards}},
  \bibinfo{author}{\bibfnamefont{J.~K.} \bibnamefont{Freericks}},
  \bibinfo{author}{\bibfnamefont{G.-D.} \bibnamefont{Lin}},
  \bibinfo{author}{\bibfnamefont{L.-M.} \bibnamefont{Duan}}, \bibnamefont{and}
  \bibinfo{author}{\bibfnamefont{C.}~\bibnamefont{Monroe}},
  \bibinfo{journal}{Nature} \textbf{\bibinfo{volume}{465}},
  \bibinfo{pages}{590} (\bibinfo{year}{2010}).

\bibitem[{\citenamefont{Johanning et~al.}(2009)\citenamefont{Johanning,
  Var\'{o}n, and Wunderlich}}]{Johanning2009}
\bibinfo{author}{\bibfnamefont{M.}~\bibnamefont{Johanning}},
  \bibinfo{author}{\bibfnamefont{A.~F.} \bibnamefont{Var\'{o}n}},
  \bibnamefont{and}
  \bibinfo{author}{\bibfnamefont{C.}~\bibnamefont{Wunderlich}},
  \bibinfo{journal}{J. Phys. B} \textbf{\bibinfo{volume}{42}},
  \bibinfo{pages}{154009} (\bibinfo{year}{2009}).

\bibitem[{\citenamefont{Bak and Bruinsma}(1982)}]{Bak1982}
\bibinfo{author}{\bibfnamefont{P.}~\bibnamefont{Bak}} \bibnamefont{and}
  \bibinfo{author}{\bibfnamefont{R.}~\bibnamefont{Bruinsma}},
  \bibinfo{journal}{Phys. Rev. Lett.} \textbf{\bibinfo{volume}{49}},
  \bibinfo{pages}{249} (\bibinfo{year}{1982}).

\bibitem[{\citenamefont{Deng et~al.}(2005)\citenamefont{Deng, Porras, and
  Cirac}}]{Deng2005}
\bibinfo{author}{\bibfnamefont{X.-L.} \bibnamefont{Deng}},
  \bibinfo{author}{\bibfnamefont{D.}~\bibnamefont{Porras}}, \bibnamefont{and}
  \bibinfo{author}{\bibfnamefont{J.~I.} \bibnamefont{Cirac}},
  \bibinfo{journal}{Phys. Rev. A} \textbf{\bibinfo{volume}{72}},
  \bibinfo{pages}{063407} (\bibinfo{year}{2005}).

\bibitem[{\citenamefont{Mermin and Wagner}(1966)}]{Mermin1966}
\bibinfo{author}{\bibfnamefont{N.~D.} \bibnamefont{Mermin}} \bibnamefont{and}
  \bibinfo{author}{\bibfnamefont{H.}~\bibnamefont{Wagner}},
  \bibinfo{journal}{Phys. Rev. Lett.} \textbf{\bibinfo{volume}{17}},
  \bibinfo{pages}{1133} (\bibinfo{year}{1966}).

\bibitem[{\citenamefont{Citro et~al.}(2007)}]{Citro2007}
\bibinfo{author}{\bibfnamefont{R.} \bibnamefont{Citro}}, \bibinfo{author}{\bibfnamefont{E.} \bibnamefont{Orignac}}, \bibinfo{author}{\bibfnamefont{S.} \bibnamefont{De~Palo}} \bibnamefont{and} \bibinfo{author}{\bibfnamefont{M.~L.} \bibnamefont{Chiofalo}},
  \bibinfo{journal}{Phys. Rev. A} \textbf{\bibinfo{volume}{75}},
  \bibinfo{pages}{051602(R)} (\bibinfo{year}{2007}).

\bibitem[{\citenamefont{Citro et~al.}(2008)}]{Citro2008}
\bibinfo{author}{\bibfnamefont{R.} \bibnamefont{Citro}}, \bibinfo{author}{\bibfnamefont{S.} \bibnamefont{De~Palo}}, \bibinfo{author}{\bibfnamefont{E.} \bibnamefont{Orignac}}, \bibinfo{author}{\bibfnamefont{P.} \bibnamefont{Pedri}} \bibnamefont{and} \bibinfo{author}{\bibfnamefont{M.~L.} \bibnamefont{Chiofalo}},
  \bibinfo{journal}{New J. Phys.} \textbf{\bibinfo{volume}{10}},
  \bibinfo{pages}{045011} (\bibinfo{year}{2008}).

\bibitem[{\citenamefont{Burnell et~al.}(2009)\citenamefont{Burnell, Parish,
  Cooper, and Sondhi}}]{Burnell2009b}
\bibinfo{author}{\bibfnamefont{F.~J.} \bibnamefont{Burnell}},
  \bibinfo{author}{\bibfnamefont{M.~M.} \bibnamefont{Parish}},
  \bibinfo{author}{\bibfnamefont{N.~R.} \bibnamefont{Cooper}},
  \bibnamefont{and} \bibinfo{author}{\bibfnamefont{S.~L.}
  \bibnamefont{Sondhi}}, \bibinfo{journal}{Phys. Rev. B}
  \textbf{\bibinfo{volume}{80}}, \bibinfo{pages}{174519}
  (\bibinfo{year}{2009}).

\bibitem[{\citenamefont{Chiaverini and Lybarger}(2008)}]{Chiaverini2008}
\bibinfo{author}{\bibfnamefont{J.}~\bibnamefont{Chiaverini}} \bibnamefont{and}
  \bibinfo{author}{\bibfnamefont{W.~E.} \bibnamefont{Lybarger},
  \bibfnamefont{Jr.}}, \bibinfo{journal}{Phys. Rev. A}
  \textbf{\bibinfo{volume}{77}}, \bibinfo{pages}{022324}
  (\bibinfo{year}{2008}).
   
\bibitem[{\citenamefont{Wessel and Troyer}(2005)}]{Wessel2005b}
\bibinfo{author}{\bibfnamefont{S.}~\bibnamefont{Wessel}} \bibnamefont{and}
  \bibinfo{author}{\bibfnamefont{M.}~\bibnamefont{Troyer}},
  \bibinfo{journal}{Phys. Rev. Lett.} \textbf{\bibinfo{volume}{95}},
  \bibinfo{pages}{127205} (\bibinfo{year}{2005}).

\bibitem[{\citenamefont{Boninsegni and Prokof'ev}(2005)}]{Boninsegni2005b}
\bibinfo{author}{\bibfnamefont{M.}~\bibnamefont{Boninsegni}} \bibnamefont{and}
  \bibinfo{author}{\bibfnamefont{N.}~\bibnamefont{Prokof'ev}},
  \bibinfo{journal}{Phys. Rev. Lett.} \textbf{\bibinfo{volume}{95}},
  \bibinfo{pages}{237204} (\bibinfo{year}{2005}).

\bibitem[{\citenamefont{Heidarian and Damle}(2005)}]{Heidarian2005}
\bibinfo{author}{\bibfnamefont{D.}~\bibnamefont{Heidarian}} \bibnamefont{and}
  \bibinfo{author}{\bibfnamefont{K.}~\bibnamefont{Damle}},
  \bibinfo{journal}{Phys. Rev. Lett.} \textbf{\bibinfo{volume}{95}},
  \bibinfo{pages}{127206} (\bibinfo{year}{2005}).

\bibitem[{\citenamefont{Melko et~al.}(2005)\citenamefont{Melko, Paramekanti,
  Burkov, Vishwanath, Sheng, and Balents}}]{Melko2005}
\bibinfo{author}{\bibfnamefont{R.~G.} \bibnamefont{Melko}},
  \bibinfo{author}{\bibfnamefont{A.}~\bibnamefont{Paramekanti}},
  \bibinfo{author}{\bibfnamefont{A.~A.} \bibnamefont{Burkov}},
  \bibinfo{author}{\bibfnamefont{A.}~\bibnamefont{Vishwanath}},
  \bibinfo{author}{\bibfnamefont{D.~N.} \bibnamefont{Sheng}}, \bibnamefont{and}
  \bibinfo{author}{\bibfnamefont{L.}~\bibnamefont{Balents}},
  \bibinfo{journal}{Phys. Rev. Lett.} \textbf{\bibinfo{volume}{95}},
  \bibinfo{pages}{127207} (\bibinfo{year}{2005}).  
  
\bibitem[{\citenamefont{Pollet et~al.}(2010)\citenamefont{Pollet, Picon,
  B\"{u}chler, and Troyer}}]{Pollet2010}
\bibinfo{author}{\bibfnamefont{L.}~\bibnamefont{Pollet}},
  \bibinfo{author}{\bibfnamefont{J.~D.} \bibnamefont{Picon}},
  \bibinfo{author}{\bibfnamefont{H.~P.} \bibnamefont{B\"{u}chler}},
  \bibnamefont{and} \bibinfo{author}{\bibfnamefont{M.}~\bibnamefont{Troyer}},
  \bibinfo{journal}{Phys. Rev. Lett.} \textbf{\bibinfo{volume}{104}},
  \bibinfo{pages}{125302} (\bibinfo{year}{2010}).    

\bibitem[{\citenamefont{Kim and Chan}(2004{\natexlab{a}})}]{Kim2004}
\bibinfo{author}{\bibfnamefont{E.}~\bibnamefont{Kim}} \bibnamefont{and}
  \bibinfo{author}{\bibfnamefont{M.~H.~W.} \bibnamefont{Chan}},
  \bibinfo{journal}{Nature} \textbf{\bibinfo{volume}{427}},
  \bibinfo{pages}{225} (\bibinfo{year}{2004}{\natexlab{a}}).

\bibitem[{\citenamefont{Kim and Chan}(2004{\natexlab{b}})}]{Kim2004a}
\bibinfo{author}{\bibfnamefont{E.}~\bibnamefont{Kim}} \bibnamefont{and}
  \bibinfo{author}{\bibfnamefont{M.~H.~W.} \bibnamefont{Chan}},
  \bibinfo{journal}{Science} \textbf{\bibinfo{volume}{305}},
  \bibinfo{pages}{1941} (\bibinfo{year}{2004}{\natexlab{b}}).

\bibitem[{\citenamefont{Rittner and Reppy}(2006)}]{Rittner2006}
\bibinfo{author}{\bibfnamefont{A.~S.~C.} \bibnamefont{Rittner}}
  \bibnamefont{and} \bibinfo{author}{\bibfnamefont{J.~D.} \bibnamefont{Reppy}},
  \bibinfo{journal}{Phys. Rev. Lett.} \textbf{\bibinfo{volume}{97}},
  \bibinfo{eid}{165301} (\bibinfo{year}{2006}).

\bibitem[{\citenamefont{Rittner and Reppy}(2007)}]{Rittner2007}
\bibinfo{author}{\bibfnamefont{A.~S.~C.} \bibnamefont{Rittner}}
  \bibnamefont{and} \bibinfo{author}{\bibfnamefont{J.~D.} \bibnamefont{Reppy}},
  \bibinfo{journal}{Phys. Rev. Lett.} \textbf{\bibinfo{volume}{98}},
  \bibinfo{eid}{175302} (\bibinfo{year}{2007}).

\bibitem[{\citenamefont{Baumann et~al.}(2010)\citenamefont{Baumann, Guerlin,
  Brennecke, and Esslinger}}]{Baumann2010}
\bibinfo{author}{\bibfnamefont{K.}~\bibnamefont{Baumann}},
  \bibinfo{author}{\bibfnamefont{C.}~\bibnamefont{Guerlin}},
  \bibinfo{author}{\bibfnamefont{F.}~\bibnamefont{Brennecke}},
  \bibnamefont{and}
  \bibinfo{author}{\bibfnamefont{T.}~\bibnamefont{Esslinger}},
  \bibinfo{journal}{Nature} \textbf{\bibinfo{volume}{464}},
  \bibinfo{pages}{1301} (\bibinfo{year}{2010}).

\bibitem[{\citenamefont{Arkhipov et~al.}(2005)\citenamefont{Arkhipov,
  Astrakharchik, Belikov, and Lozovik}}]{Arkhipov2005}
\bibinfo{author}{\bibfnamefont{A.~S.} \bibnamefont{Arkhipov}},
  \bibinfo{author}{\bibfnamefont{G.~E.} \bibnamefont{Astrakharchik}},
  \bibinfo{author}{\bibfnamefont{A.~V.} \bibnamefont{Belikov}},
  \bibnamefont{and} \bibinfo{author}{\bibfnamefont{Y.~E.}
  \bibnamefont{Lozovik}}, \bibinfo{journal}{JETP Lett.}
  \textbf{\bibinfo{volume}{82}}, \bibinfo{pages}{39} (\bibinfo{year}{2005}).

\bibitem[{\citenamefont{B\"{u}chler et~al.}(2007)\citenamefont{B\"{u}chler,
  Demler, Lukin, Micheli, Prokof'ev, Pupillo, and Zoller}}]{Buechler2007}
\bibinfo{author}{\bibfnamefont{H.~P.} \bibnamefont{B\"{u}chler}},
  \bibinfo{author}{\bibfnamefont{E.}~\bibnamefont{Demler}},
  \bibinfo{author}{\bibfnamefont{M.}~\bibnamefont{Lukin}},
  \bibinfo{author}{\bibfnamefont{A.}~\bibnamefont{Micheli}},
  \bibinfo{author}{\bibfnamefont{N.}~\bibnamefont{Prokof'ev}},
  \bibinfo{author}{\bibfnamefont{G.}~\bibnamefont{Pupillo}}, \bibnamefont{and}
  \bibinfo{author}{\bibfnamefont{P.}~\bibnamefont{Zoller}},
  \bibinfo{journal}{Phys. Rev. Lett.} \textbf{\bibinfo{volume}{98}},
  \bibinfo{pages}{060404} (\bibinfo{year}{2007}).

\bibitem[{\citenamefont{Lindemann}(1910)}]{Lindemann1910}
\bibinfo{author}{\bibfnamefont{F.~A.} \bibnamefont{Lindemann}},
  \bibinfo{journal}{Z. Phys.} \textbf{\bibinfo{volume}{11}},
  \bibinfo{pages}{609} (\bibinfo{year}{1910}).

\bibitem[{\citenamefont{Albuquerque et~al.}(2007)\citenamefont{Albuquerque,
  Alet, Corboz, Dayal, Feiguin, Fuchs, Gamper, Gull, G\"{u}rtler, Honecker
  et~al.}}]{alps}
\bibinfo{author}{\bibfnamefont{A.}~\bibnamefont{Albuquerque}},
  \bibinfo{author}{\bibfnamefont{F.}~\bibnamefont{Alet}},
  \bibinfo{author}{\bibfnamefont{P.}~\bibnamefont{Corboz}},
  \bibinfo{author}{\bibfnamefont{P.}~\bibnamefont{Dayal}},
  \bibinfo{author}{\bibfnamefont{A.}~\bibnamefont{Feiguin}},
  \bibinfo{author}{\bibfnamefont{S.}~\bibnamefont{Fuchs}},
  \bibinfo{author}{\bibfnamefont{L.}~\bibnamefont{Gamper}},
  \bibinfo{author}{\bibfnamefont{E.}~\bibnamefont{Gull}},
  \bibinfo{author}{\bibfnamefont{S.}~\bibnamefont{G\"{u}rtler}},
  \bibinfo{author}{\bibfnamefont{A.}~\bibnamefont{Honecker}},
  \bibnamefont{et~al.}, \bibinfo{journal}{J. Magn. Magn. Mater.}
  \textbf{\bibinfo{volume}{310}}, \bibinfo{pages}{1187} (\bibinfo{year}{2007}).

\bibitem[{\citenamefont{White}(1992)}]{White1992b}
\bibinfo{author}{\bibfnamefont{S.~R.} \bibnamefont{White}},
  \bibinfo{journal}{Phys. Rev. Lett.} \textbf{\bibinfo{volume}{69}},
  \bibinfo{pages}{2863} (\bibinfo{year}{1992}).

\bibitem[{\citenamefont{Schollw{\"o}ck}(2005)}]{Schollwock2005}
\bibinfo{author}{\bibfnamefont{U.}~\bibnamefont{Schollw{\"o}ck}},
  \bibinfo{journal}{Rev. Mod. Phys.} \textbf{\bibinfo{volume}{77}},
  \bibinfo{pages}{259} (\bibinfo{year}{2005}).

\bibitem[{\citenamefont{Eckardt}()}]{Andreprivate}
\bibinfo{author}{\bibfnamefont{A.}~\bibnamefont{Eckardt}},
  \bibinfo{note}{private communication} (\bibinfo{year}{2010}).

\bibitem[{\citenamefont{Vidal}(2007)}]{Vidal2007}
\bibinfo{author}{\bibfnamefont{G.}~\bibnamefont{Vidal}},
  \bibinfo{journal}{Phys. Rev. Lett.} \textbf{\bibinfo{volume}{98}},
  \bibinfo{pages}{070201} (\bibinfo{year}{2007}).

\bibitem[{\citenamefont{Murg et~al.}(2008)\citenamefont{Murg, Cirac, Pirvu, and
  Verstraete}}]{Murg2008a}
\bibinfo{author}{\bibfnamefont{V.}~\bibnamefont{Murg}},
  \bibinfo{author}{\bibfnamefont{J.~I.} \bibnamefont{Cirac}},
  \bibinfo{author}{\bibfnamefont{B.}~\bibnamefont{Pirvu}}, \bibnamefont{and}
  \bibinfo{author}{\bibfnamefont{F.}~\bibnamefont{Verstraete}}
  (\bibinfo{year}{2008}), \eprint{arXiv:0804.3976}.

\bibitem[{\citenamefont{Chen et~al.}(2008)\citenamefont{Chen, Wang, Hao, and
  Wang}}]{Chen2008}
\bibinfo{author}{\bibfnamefont{S.}~\bibnamefont{Chen}},
  \bibinfo{author}{\bibfnamefont{L.}~\bibnamefont{Wang}},
  \bibinfo{author}{\bibfnamefont{Y.}~\bibnamefont{Hao}}, \bibnamefont{and}
  \bibinfo{author}{\bibfnamefont{Y.}~\bibnamefont{Wang}},
  \bibinfo{journal}{Phys. Rev. A} \textbf{\bibinfo{volume}{77}},
  \bibinfo{pages}{032111} (\bibinfo{year}{2008}).

\end{thebibliography}

\end{document}